\documentclass[letter]{aa} 

\usepackage{graphicx}
\usepackage{lscape}
\usepackage{txfonts}
\usepackage{color}
\begin{document} 

\title{Discovery of the elusive radical NCO and confirmation of H$_2$NCO$^+$ in space
\thanks{Based on observations carried out with the IRAM 30m telescope. 
IRAM is supported by INSU/CNRS (France), MPG (Germany) and IGN (Spain).}
}

\author{
N. Marcelino \inst{1}
\and
M. Ag\'undez \inst{1}
\and
J. Cernicharo \inst{1}
\and
E. Roueff \inst{2}
\and
M. Tafalla  \inst{3}
}

\institute{
Instituto de F\'isica Fundamental, CSIC, C/ Serrano 123, 28006 Madrid, Spain
\and
Sorbonne Universit\'e, Observatoire de Paris, Universit\'e PSL, CNRS, LERMA, F-92190, Meudon, France
\and
Observatorio Astron\'omico Nacional (OAN), C/ Alfonso XII 3, 28014 Madrid, Spain
}

\date{Received ; accepted }


\abstract{
The isocyanate radical (NCO) is the simplest molecule containing the backbone of the 
peptide bond, C(=O)-N. This bond has a prebiotic interest since is the one linking two 
amino acids to form large chains of proteins. It is also present in some organic molecules 
observed in space such as HNCO, NH$_2$CHO and CH$_3$NCO. 
In this letter we report the first detection in space of NCO towards the dense core L483. 
We also report the identification of the ion H$_2$NCO$^+$, definitively confirming its 
presence in space, and observations of HNCO, HOCN, and HCNO in the same source.
For NCO, we derive a column density of $2.2\times10^{12}$ cm$^{-2}$, which means that it 
is only $\sim$5 times less abundant than HNCO. We find that H$_2$NCO$^+$, HOCN and HCNO 
have abundances relative to HNCO of 1/400, 1/80, and 1/160, respectively.
Both NCO and H$_2$NCO$^+$ are involved in the production of HNCO and several of its isomers. 
We have updated our previous chemical models involving NCO and the production of the CHNO isomers. 
Taking into account the uncertainties in the model, the observed abundances are reproduced relatively well. 
Indeed, the detection of NCO and H$_2$NCO$^+$ in L483 supports the chemical pathways to the formation 
of the detected CHNO isomers. Sensitive observations of NCO in sources where other molecules 
containing the C(=O)-N subunit have been detected could help in elucidating its role in prebiotic chemistry 
in space.}

\keywords{Astrochemistry --
                ISM: clouds, L483 --
                ISM: abundances --
                Stars: formation, low-mass
                Line: identification
               }


\maketitle
%

\section{Introduction}

Most molecules observed in space can be formed with just four atoms, H, C, N, and O. These atoms are the building blocks of organic and prebiotic molecules, and arranged together, constitute the backbone of the peptide bond R--C(=O)--N(--H)--R$'$, which links two amino acids and allows to build large proteins. Therefore, the observation of simple molecules with the C(=O)--N group in space can provide important clues on the earliest chemical steps in the synthesis of amino acids. The isocyanate radical (NCO) is the simplest such species and it is predicted to be abundant in dark clouds \citep{prasad78,marce09}. It is also the main precursor of isocyanic acid (HNCO), a species that has been found in a large variety of interstellar environments \citep[see][and references within]{marce09}.

HNCO has several metastable isomers: HOCN, HCNO, and HONC, which lie respectively at 24.7, 70.7, and 84.1 kcal mol$^{-1}$ respect to HNCO \citep[see][]{schuurman04}. Two of these isomers, HCNO and HOCN, have also been previously detected in molecular clouds \citep{marce09,marce10,brunken10}. The chemical pathways to the formation of these species are diverse, but the main precursors in gas-phase are the radicals NCO and CNO \citep{marce10,quan10}. Followed by protonation and hydrogenation, these radicals will lead to the different protonated CHNO ions: H$_2$NCO$^+$, HNCOH$^+$, HCNOH$^+$, H$_2$OCN$^+$, and H$_2$CNO$^+$ \citep{marce10,quan10}. By dissociative recombination reactions, these ions could produce all the CHNO isomers, including 
the highest isomer in energy HONC not detected so far in space, and also the NCO radical \citep{marce10}. 
The two isomeric forms of the protonated isocyanic acid H$_2$NCO$^+$ and HNCOH$^+$, have been studied in the laboratory \citep{lattanzi12,gupta13}, and H$_2$NCO$^+$ has been tentatively detected in absorption of the low-lying rotational transitions in the centimeter band towards the molecular cloud Sgr B2(N) by \citet{gupta13}. 

\begin{figure*}
\centering
\includegraphics[width=0.9\hsize]{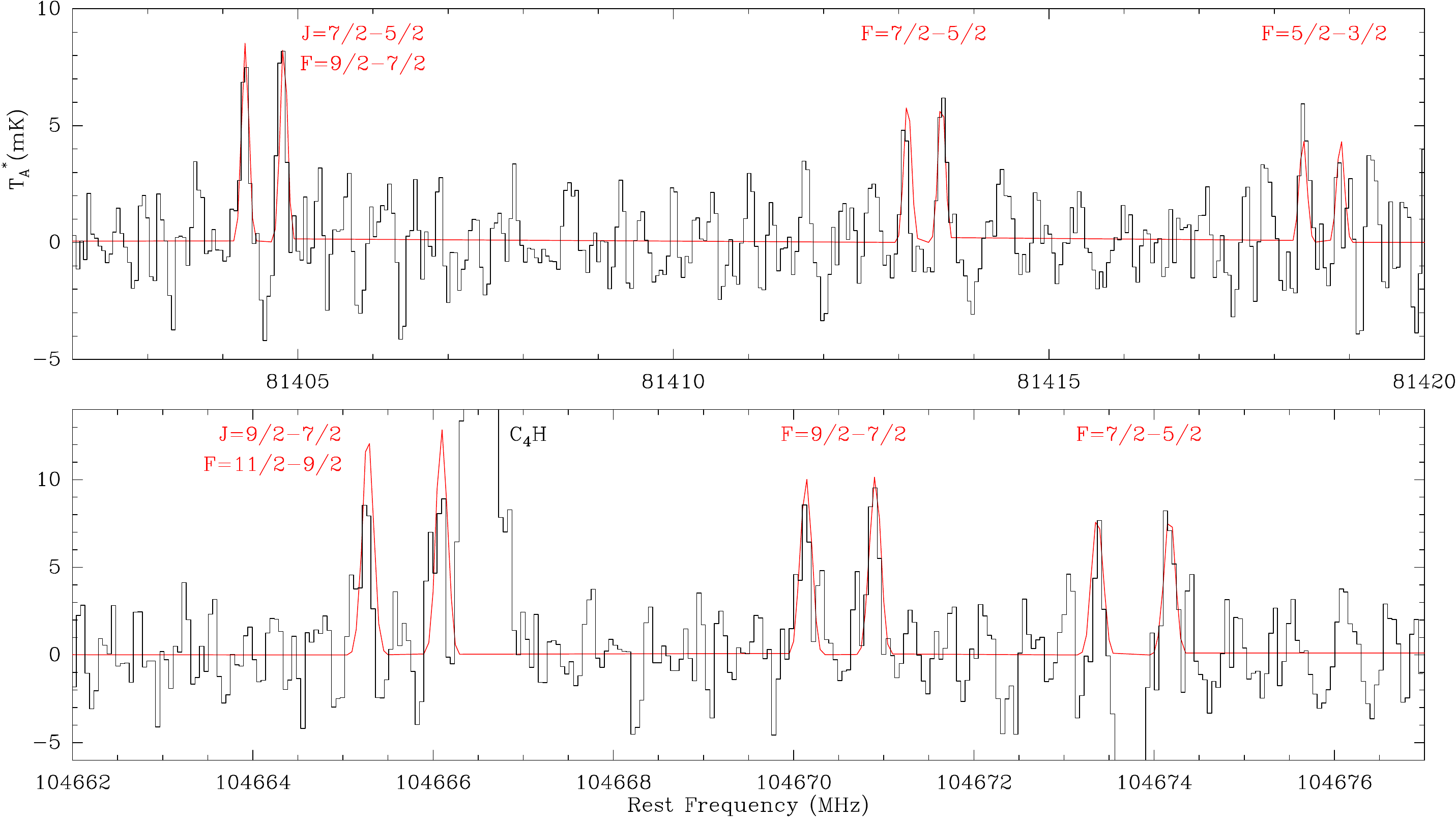}
\caption{Observed transitions of NCO in the $^2\Pi_{3/2}$ state towards L483. The strongest components of the $J=7/2-5/2$ and $J=9/2-7/2$ transitions are shown in the upper and lower panels, respectively. LTE results are overplotted in red. 
Note that the absorption feature in the lower panel between the doublet $F=7/2-5/2$ is not real, it is the corresponding negative of the C$_4$H line due to the Frequency Switching observing mode (located 7.2 MHz away).}
\label{fig:fig-nco}
\end{figure*}

In this letter, we report the first detection in space of NCO and we confirm the presence of H$_2$NCO$^+$ through millimeter emission transitions. These species were detected towards L483, a dense core located in the Aquila Rift which hosts a protostar, IRAS\,18148$-$0440, in transition from Class 0 to Class I and which shows infall motions and a collimated molecular outflow \citep{tafalla00,park00}. Based on the bright emission of carbon chains like C$_4$H \citep{agundez08,sakai09}, it has been suggested that L483 may host a warm-carbon-chain-chemistry environment. Apart from carbon chains, L483 is also rich in O-bearing organic molecules such as HCO, HCCO, H$_2$CCO, CH$_3$CHO, HCCCHO, and $c$-C$_3$H$_2$O \citep{agundez15a,loison16}. Recently, the chemical richness of the source has been evidenced with the discovery of several new molecules: HCCO, NCCNH$^+$, NS$^+$, HCS, and HSC \citep{agundez15a,agundez15b,agundez18,cerni18}. Recent ALMA observations have shown a chemical differentiation in L483, with carbon chains such as C$_2$H tracing the envelope and more complex organics like HNCO, NH$_2$CHO, and HCOOCH$_3$ being distributed around the protostar \citep{oya17}. The detection of NCO and H$_2$NCO$^+$ around a low-mass protostar such as L483 can thus shed light on the formation of prebiotic molecules containing the C(=O)--N group, like HNCO and its isomers, NH$_2$CHO, and CH$_3$NCO \citep{cerni16,quenard18}.

\section{Observations}

The data presented here are part of a full 3\,mm (80-116 GHz) spectral line survey of L483, using the IRAM 30m radiotelescope in Granada (Spain). The line survey observations were performed in several sessions, in August and November 2016, and in May and December 2017. We also obtained Director's Discretionary Time to confirm the detection of NCO and H$_2$NCO$^+$ in January 2018. 
The observed position corresponds to that of the infrared source IRAS\,18148--0440, $\alpha_{J2000}=18^{\rm h} 17^{\rm m} 29.8^{\rm s}$ and $\delta_{J2000}=-04^\circ 39' 38.0''$ \citep{fuller93}.

We used the EMIR receivers operating at 3\,mm, connected to the Fast Fourier Transform Spectrometers (FTS) in high resolution mode, providing a spectral resolution of 50 kHz, which corresponds to velocity resolutions of $\sim$0.13--0.19 km\,s$^{-1}$ in the 80-116 GHz range. 
Thanks to the versatility of the EMIR and FTS, we could observe 4 different bands per spectral setup covering 4$\times$1.8 GHz of bandwidth. All the observations were done in Frequency Switching mode, with a frequency throw of 7.2 MHz. Pointing and Focus were checked every 1 and 3 hours, respectively. Pointing errors were always within 3$''$. The 30\,m beam sizes at 3\,mm are between 30$''$ and 21$''$. Weather conditions were different from one observing period to other, ranging from good winter conditions, with 2-5 mm of precipitable water vapor (pwv), to average summer conditions, with pwv$\leq$10 mm. Concerning the data presented here, system temperatures range between 90 and 120 K and the final rms noise is $\sim$1-4 mK, depending on frequency. The spectra were calibrated in antenna temperature corrected for atmospheric attenuation and for antenna ohmic and spillover losses using the ATM package \citep{cerni85,pardo01}. Data reduction and analysis were done using the CLASS program of the GILDAS software\footnote{http://www.iram.fr/IRAMFR/GILDAS}. 

\section{Results}

During the analysis of the spectral survey data, we have found a number of unidentified lines. Among them there are several features at 81.4 GHz and 104.6 GHz which are coincident with the strongest components of the $J= 7/2-5/2$ and $J= 9/2-7/2$ transitions of the NCO radical in the $^2\Pi_{3/2}$ 
state (see Figure~\ref{fig:fig-nco}). The radical NCO is a linear triatomic molecule with a $^2\Pi_i$ electronic ground state, where the $^2\Pi_{1/2}$ spin-orbit state lies 137 K above the $^2\Pi_{3/2}$ state. The rotational levels are split by $\Lambda$ doubling, leading to $e$ and $f$ parities, and are further split due to the interaction with the nuclear spin of $^{14}$N, leading to a hyperfine structure (\citealt{saito70,kawaguchi85}; see also CDMS\footnote{\texttt{http://www.astro.uni-koeln.de/cdms/}}; \citealt{muller05}, and MADEX\footnote{\texttt{https://nanocosmos.iff.csic.es/$?$page\_id=1619}}; \citealt{madex}). Already in 1978, this radical was predicted to be abundant in dark clouds \citep{prasad78}, although it has not been yet detected in space probably because of its low dipole moment \citep[0.64 D,][]{saito70}. The complex rotational spectrum of NCO helps in the assignment of the lines. The strongest components of the $J=7/2-5/2$ and $J=9/2-7/2$ are detected in L483, with no missing line, and with center frequencies and relative intensities in very good agreement with the values obtained from laboratory. With the identification of 12 different transitions, the detection of NCO can be considered secure. 

\begin{figure}
\includegraphics[width=0.9\hsize]{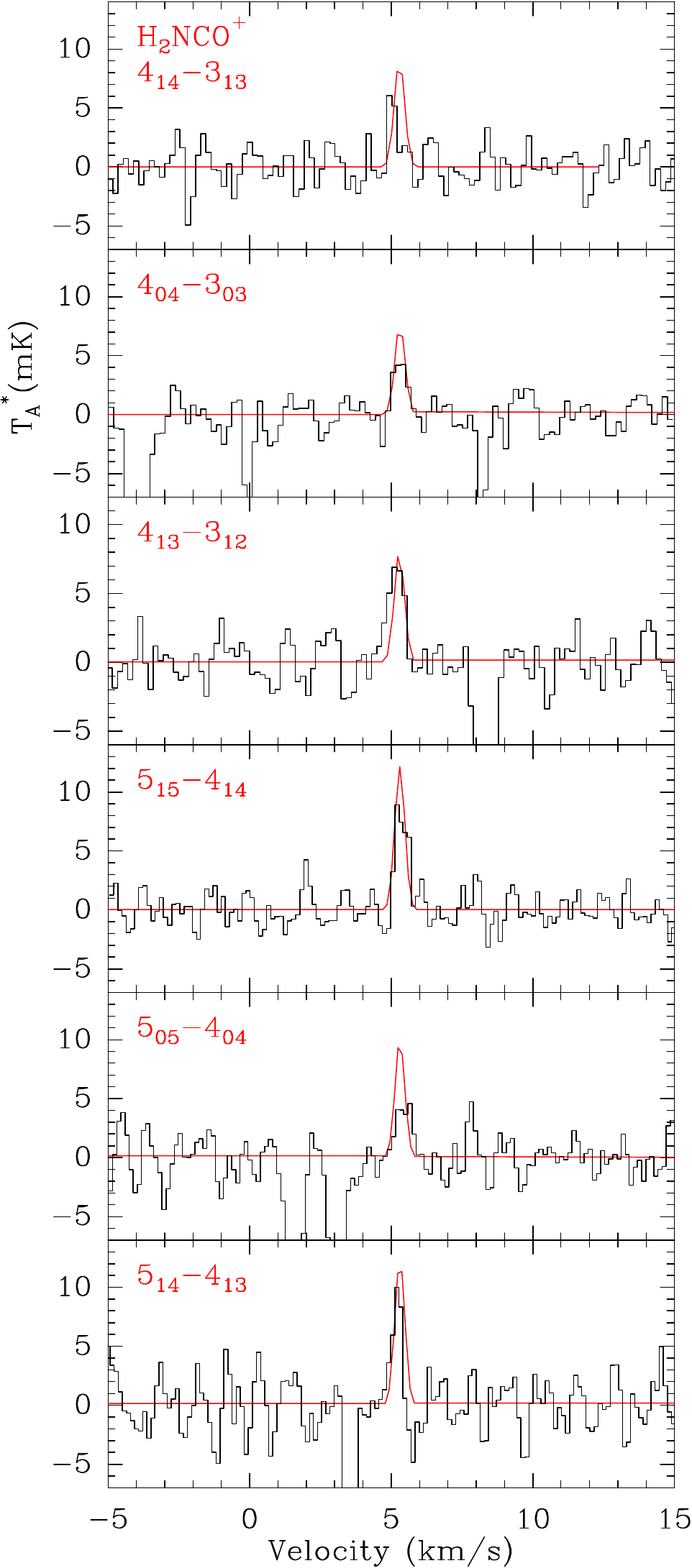}
\caption{Observed transitions of H$_2$NCO$^+$ towards L483. LTE results are overplotted in red.}
\label{fig:fig-h2ncop}
\end{figure}

We have also detected six lines arising from H$_2$NCO$^+$, the lowest energy isomer of protonated isocyanic acid (see Figure~\ref{fig:fig-h2ncop}). This ion has a large dipole moment \citep[4.13 D,][]{lattanzi12} and has been tentatively detected towards the molecular clouds in the Galactic Center Sgr B2(N) in absorption at 20 and 40 GHz by \citet{gupta13}. H$_2$NCO$^+$ is a planar molecular ion with a C$_{2\rm{v}}$ symmetry, in which the two H nuclei are equivalent. Thus, the rotational levels are separated into {\it ortho} ($K_a$ odd) and {\it para} ($K_a$ even) states, with statistical weights 3:1. Due to the $^{14}$N nuclear spin, H$_2$NCO$^+$ also presents quadrupole hyperfine structure. However it is unresolved in the L483 spectra (see Fig.~\ref{fig:fig-h2ncop}) and we do not take it into account in the analysis below. Although the lines of H$_2$NCO$^+$ in L483 are weak, with $T_A^*$ between 5 and 10 mK, all of them are detected above the $3\sigma$ level (rms noise levels are 1.2-2 mK). Furthermore, the fact that all the strongest transitions covered in the 3 mm band are detected, well centered at the systemic velocity of L483 ($V_{LSR}$ $\sim$ 5.3 km s$^{-1}$; \citealt{fuller93,agundez08}), supports the identification.

Table~\ref{table:lines} lists all the observed transitions, together with the line parameters derived from gaussian fits. 
Rest frequencies and spectroscopic data were taken from the MADEX catalogue \citep{madex}. We have computed column densities assuming Local Thermodynamic Equilibrium (LTE) and $T_{\rm rot}=10$ K, which is consistent with the values obtained for other species in this cloud \citep{agundez15a,agundez15b}. For NCO, since the range of upper level energies covered by the observed transitions is narrow (6.6-11.7 K; see Table~\ref{table:lines}), the rotational temperature is poorly constrained using rotational diagrams. In the case of H$_2$NCO$^+$, we have a relatively low number of transitions available when considering {\it ortho} and {\it para} species separately (in particular only two transitions are observed for the {\it para} state). Derived column densities are $N=(2.2\pm0.7)\times10^{12}$ cm$^{-2}$ for NCO, and $3.0\times10^{10}$ cm$^{-2}$ for H$_2$NCO$^+$ 
(values for the {\it ortho} and {\it para} species are $(2.1\pm0.5)\times10^{10}$ cm$^{-2}$ and $(8.6\pm3.9)\times10^{9}$ cm$^{-2}$, respectively).
In the line survey we have also covered two transitions of HNCO, HOCN and HCNO. We have computed column densities for all these species assuming $T_{\rm rot}=10$ K. The fourth isomer, HONC, and the ion HNCOH$^+$ are not detected, and hence only upper limits are provided.
The column densities and fractional abundances derived are listed for all species in Table~\ref{table:results}. We used $N({\rm H_2})=3\times10^{22}$ cm$^{-2}$, which was derived by \citet{tafalla00} from observations of C$^{17}$O. 

\begin{table}
\caption{Column densities and fractional abundances derived.}
\label{table:results}
\centering
\begin{tabular}{lcrr@{\hspace{0.05cm}}r}
\hline\hline
Species       &  $N$ (cm$^{-2}$)  & \multicolumn{1}{c}{$X$/H$_2$} & \multicolumn{1}{c}{HNCO/$X$} \\
\hline
NCO           &  $2.2\times10^{12}$    &  $7.3\times10^{-11}$    & 5.5 \\
H$_2$NCO$^+$  &  $3.0\times10^{10}$    &  $1.0\times10^{-12}$    & 400 \\
HNCOH$^+$     &  $<$$2.3\times10^{10}$ &  $<$$7.7\times10^{-13}$ & $>$520 \\
HNCO          &  $1.2\times10^{13}$    &  $4.1\times10^{-10}$    & -- \\
HOCN          &  $1.5\times10^{11}$    &  $5.0\times10^{-12}$    & 80 \\
HCNO          &  $7.3\times10^{10}$    &  $2.4\times10^{-12}$    & 160 \\
HONC          &  $<$$8.3\times10^{9}$  &  $<$$2.8\times10^{-13}$ & $>$1450 \\
\hline
\end{tabular}
\end{table}

\section{Discussion}

The chemical processes linking H, C, O and N atoms take place in combustion processes but the kinetics at the low temperatures of interstellar clouds is subject to many uncertainties. To investigate the formation of NCO and H$_2$NCO$^+$ in L483, we have updated the chemical schema used in \citet{marce09,marce10} by considering the UMIST \citep{mcelroy13} and KIDA \citep{wakelam15} databases. In the absence of experimental or theoretical information on the kinetics, we have verified the energy balance of various reactions. Table~\ref{tab:enthalpy} reports our estimate of the formation enthalpies of different isomers of the CHNO$^+$ and CH$_2$NO$^+$ ions, which are not all reported in thermodynamic tables nor in the KIDA database. We find that some reactions suggested in KIDA are in fact endothermic, e.g.,
\begin{equation}
{\rm HOCN^+   + H_2    \rightarrow   H_2OCN^+ +  H,}
\end{equation}
\begin{equation}
{\rm HCNO^+   + H_2    \rightarrow   H_2CNO^+ + H.}
\end{equation}
It is obvious that our tentative chemical network is subject to many uncertainties. Amongst these, the dissociative recombination (DR) rate coefficients of the CH$_2$NO$^+$ ions and the branching ratios are quite critical. The present model results
are obtained when ejection of H and H$_2$ are favoured compared to other neutral channel products. In the cases where two CHNO isomers are produced in the DR, we have also assumed that the most stable is produced preferentially. The corresponding reaction rate coefficients are displayed in Table~\ref{tab:DRrates}.
Another uncertainty in our chemical network is the formation route of H$_2$NCO$^+$ through the NH$_3$ + HCO$^+$ reaction, 
which is exothermic. 
For this channel we assume a rate coefficient of 2.5$\times$10$^{-10}$(T/300)$^{-0.5}$ cm$^3$ s$^{-1}$, while for the rapid proton transfer channel leading to NH$_4^+$ + CO we consider a higher rate coefficient of 2.2$\times$10$^{-9}$(T/300)$^{-0.5}$ cm$^3$ s$^{-1}$.

\begin{figure}
\includegraphics[width=1.02\columnwidth]{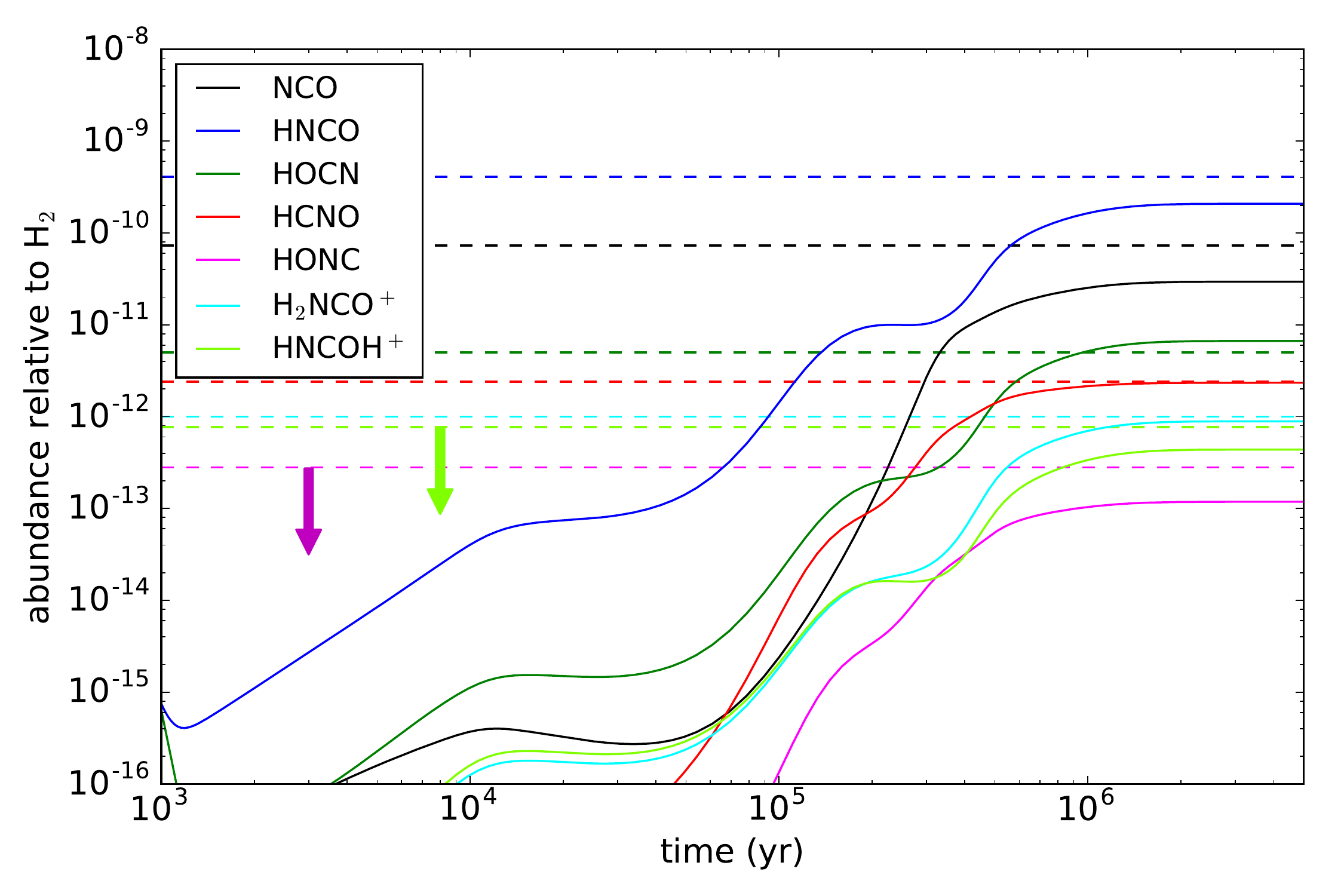}
\caption{Fractional abundances calculated as a function of time with our chemical model (see text). Horizontal dashed lines indicate the abundances derived from observations.}
\label{fig:model}
\end{figure}

In Fig.~\ref{fig:model} we show the time evolution of the abundances of various species of interest, as given by our gas-phase chemical model. We have adopted a H$_2$ volume density of $3.4\times10^4$ cm$^{-3}$ and a gas kinetic temperature of 10 K, which are adequate for the L483 cloud \citep{fuller93,anglada97,jorgensen02}. The cosmic ionization rate is fixed to $1.3\times10^{-17}$ s$^{-1}$ and the elemental abundances of C, N, O, S relative to hydrogen are set to $9\times10^{-6}$, $1\times10^{-5}$, $1.5\times10^{-5}$ and $2\times10^{-8}$, respectively. The relative abundances for the detected species are rather well reproduced by our model, while those of HONC and HNCOH$^+$ are below the observed upper limits. Given the uncertainties of the chemical network, the agreement between the model and the observations is quite satisfactory.

The different species treated in this paper appear to follow different reaction routes, both through neutral-neutral reactions and ion-molecule schema. NCO is formed straightforwardly through the CN + O$_2$ reaction at a reaction rate of $2.4\times10^{-11}$ cm$^3$ s$^{-1}$ \citep{glarborg98}. Fulminic acid (HCNO), despite at an energy of about 68 kcal mol$^{-1}$ above the most stable isomer HNCO, is one of the main products, together with HCN, of the reaction between CH$_2$ and NO, as found in the laboratory \citep{grussdorf94} and explained theoretically \citep{roggenbuck98}. HNCO and HOCN, on the other hand, are rather formed through ion molecule schema as products in the dissociative recombination of H$_2$NCO$^+$ and HNCOH$^+$, although the reaction for the latter has not yet been studied in the laboratory. The detection of H$_2$NCO$^+$ validates the occurrence of the ion-molecule formation schema.

\section{Conclusions}

In this letter we report the first detection in space of the NCO radical with a significant abundance, only $\sim$5 times lower than HNCO, towards the low-mass protostar L483. Together with H$_2$NCO$^+$, detected as well in this dense core, they are involved in the production of HNCO and some of its isomers. In L483 we obtain similar abundances of HOCN and HCNO, which is consistent with what was found by \citet{marce10} in dense cores. It is possible that NCO and H$_2$NCO$^+$ could be present in these sources as well. Unfortunately the corresponding frequencies were either not covered or the available data were not sensitive enough. Observations of NCO in sources at different stages across the star formation process could help to dilucidate its role in the synthesis of CHNO isomers and prebiotic molecules containing the C(=O)--N group.

\begin{acknowledgements}

We thank the referee for his/her useful comments, and the IRAM 30m staff for their help during the observations and time allocation of the DDT project. We acknowledge funding support from the European Research Council (ERC Grant 610256: NANOCOSMOS) and from Spanish MINECO through grant AYA2016-75066-C2-1-P. M.A. also acknowledges funding support from the Ram\'on y Cajal programme of Spanish MINECO (RyC-2014-16277). E.R. acknowledges partial support by the French program “Physique et Chimie du Milieu Interstellaire” (PCMI) funded by the Conseil National de la Recherche Scientifique (CNRS) and Centre National d’Etudes Spatiales (CNES). M.T. acknowledges support from MINECO through grant AYA2016-79006-P.

\end{acknowledgements}

%
%

\clearpage

\begin{appendix} 

\section{Additional tables}

\begin{table*}
\caption{Observed transitions and line parameters derived from Gaussian fits.
Rest frequencies and spectroscopic data are taken from the MADEX catalogue \citep{madex}.}
\label{table:lines}
\centering
\begin{tabular}{lrcccccc}
\hline\hline
Transition  & Frequency & $E_{up}$ & $A_{ul}$   & $\int T_{\rm A}^* dv$ & $V_{\rm LSR}$ &   $\Delta$v   & $T_{\rm A}^*$ \\
                  & (MHz)        & (K)         & (s$^{-1}$) &  (mK km s$^{-1}$)       & (km s$^{-1}$)   & (km s$^{-1}$) &    (mK)          \\
\hline
NCO ($^2\Pi_{3/2}$) \\
\hline
$J=7/2-5/2~~~F=9/2-7/2~~~~e$  &   81404.300 &  6.6 & 9.19 10$^{-7}$ & 3.6(9) & 5.32( 4) & 0.42(10) &  8.0 \\
$J=7/2-5/2~~~F=9/2-7/2~~~~f$   &   81404.813 &  6.6 & 9.19 10$^{-7}$ & 4.1(9) & 5.41( 3) & 0.42( 7) & 9.0 \\
$J=7/2-5/2~~~F=7/2-5/2~~~~e$  &   81413.120 &  6.6 & 8.44 10$^{-7}$ & 2.1(9) & 5.45( 4) & 0.34(10) &  5.9 \\
$J=7/2-5/2~~~F=7/2-5/2~~~~f$   &   81413.573 &  6.6 & 8.44 10$^{-7}$ & 3.2(9) & 5.26( 4) & 0.45(10) &  6.6 \\
$J=7/2-5/2~~~F=5/2-3/2~~~~e$  &   81418.385 &  6.6 & 8.17 10$^{-7}$ & 2.9(9) & 5.24( 6) & 0.44(11) &  6.2 \\
$J=7/2-5/2~~~F=5/2-3/2~~~~f$   &   81418.884 &  6.6 & 8.17 10$^{-7}$ & 1.6(9) & 5.43( 5) & 0.27(17) &  5.4 \\
$J=9/2-7/2~~~F=11/2-9/2~~e$ & 104665.278 & 11.7 & 2.19 10$^{-6}$ & 5.0(9) & 5.45( 5) & 0.55(10) &  8.6 \\
$J=9/2-7/2~~~F=11/2-9/2~~f$  & 104666.098 & 11.7 & 2.19 10$^{-6}$ & 5.4(9) & 5.43( 4) & 0.55( 7) &  9.1 \\
$J=9/2-7/2~~~F=9/2-7/2~~~~e$   & 104670.139 & 11.7 & 2.08 10$^{-6}$ & 3.6(9) & 5.36( 4) & 0.45(11) &  7.6 \\
$J=9/2-7/2~~~F=9/2-7/2~~~~f$    & 104670.905 & 11.7 & 2.08 10$^{-6}$ & 3.9(9) & 5.35( 3) & 0.37( 8) &  9.9 \\
$J=9/2-7/2~~~F=7/2-5/2~~~~e$   & 104673.371 & 11.7 & 2.05 10$^{-6}$ & 1.6(9) & 5.26( 4) & 0.22( 7) &  7.1 \\
$J=9/2-7/2~~~F=7/2-5/2~~~~f$    & 104674.173 & 11.7 & 2.05 10$^{-6}$ & 2.9(9) & 5.32( 4) & 0.35( 8) &  7.8 \\
\hline
H$_2$NCO$^+$ \\
\hline
$4_{1,4} - 3_{1,3}$  &  	80246.376 &  8.7 & 4.28 10$^{-5}$ & 2.0(5) & 5.02( 1) & 0.18(60) & 10.4 \\
$4_{0,4} - 3_{0,3}$  & 	80906.926 &  9.7 & 4.67 10$^{-5}$ & 2.8(5) & 5.32( 5) & 0.53( 9) &  4.9 \\
$4_{1,3} - 3_{1,2}$  & 	81565.636 &  8.8 & 4.49 10$^{-5}$ & 4.9(9) & 5.16( 5) & 0.63(10) &  7.3 \\
$5_{1,5} - 4_{1,4}$  & 100306.949 & 13.5 & 8.74 10$^{-5}$ & 5.0(5) & 5.37( 3) & 0.53( 5) &  8.9 \\
$5_{0,5} - 4_{0,4}$  & 101131.130 & 14.6 & 9.33 10$^{-5}$ & 3.8(9) & 5.45( 7) & 0.70(15) &  5.0 \\
$5_{1,4} - 4_{1,3}$  & 101955.974 & 13.7 & 9.18 10$^{-5}$ & 4.2(9) & 5.20( 4) & 0.38( 8) & 10.4 \\
\hline
HNCOH$^+$ \\
\hline
$5 - 4$  &  99559.525 & 14.3 & 8.82 10$^{-6}$ & $<$0.8 & & & $<$3.3 \\
\hline
HNCO \\
\hline
$4_{0,4}-3_{0,3}~~~F=3-3$  &  87924.381 & 10.5 & 7.25 10$^{-7}$ &  19(1) & 5.33(1) & 0.33(2) & 53.5 \\
$4_{0,4}-3_{0,3}~~~F=5-4$  &  87925.252 & 10.5 & 9.02 10$^{-6}$ & 761(2) & 5.31(1) & 0.44(1) & 1619.6 \\
$4_{0,4}-3_{0,3}~~~F=4-3$  &  87925.252 & 10.5 & 8.46 10$^{-6}$ &  \\
$4_{0,4}-3_{0,3}~~~F=3-2$  &  87925.252 & 10.5 & 8.29 10$^{-6}$ &  \\
$4_{0,4}-3_{0,3}~~~F=4-4$  &  87925.898 & 10.5 & 5.64 10$^{-7}$ &  24(1) & 5.30(1) & 0.40(2) & 57.4 \\
$5_{0,5}-4_{0,4}~~~F=4-4$  & 109904.922 & 15.8 & 8.81 10$^{-7}$ &  12(1) & 5.33(1) & 0.34(3) & 33.2 \\
$5_{0,5}-4_{0,4}~~~F=6-5$  & 109905.758 & 15.8 & 1.80 10$^{-5}$ & 553(1) & 5.31(1) & 0.36(1) & 1450.3 \\
$5_{0,5}-4_{0,4}~~~F=5-4$  & 109905.758 & 15.8 & 1.73 10$^{-5}$ &  \\
$5_{0,5}-4_{0,4}~~~F=4-3$  & 109905.758 & 15.8 & 1.71 10$^{-5}$ &  \\
$5_{0,5}-4_{0,4}~~~F=5-5$  & 109906.430 & 15.8 & 7.21 10$^{-7}$ &  11(1) & 5.31(2) & 0.37(4) & 28.2 \\
\hline
HOCN \\
\hline
$4_{0,4}-3_{0,3}$  &  83900.569 & 10.1 & 4.18 10$^{-5}$ & 50(1) & 5.31(1) & 0.47(1) & 99.7 \\
$5_{0,5}-4_{0,4}$  & 104874.676 & 15.1 & 8.36 10$^{-5}$ & 34(1) & 5.30(1) & 0.38(1) & 83.2 \\
\hline
HCNO \\
\hline
$4 - 3$  &  91751.320 & 11.0 & 3.84 10$^{-5}$ & 21(1) & 5.33(1) & 0.41(2) & 47.4 \\
$5 - 4$  & 114688.383 & 16.5 & 7.67 10$^{-5}$ &  8(2) & 5.35(3) & 0.26(5) & 28.6 \\
\hline
HONC \\
\hline
$4 - 3$  &  87625.193 & 10.5 & 3.41 10$^{-5}$ & $<$2.1 & & & $<$8.1 \\
$5 - 4$  & 109530.044 & 15.8 & 6.81 10$^{-5}$ & $<$1.6 & & & $<$6.9 \\
\hline
\end{tabular}
\end{table*}

\begin{table}
\caption{Formation enthalpy estimates in kcal mol$^{-1}$}             
\label{tab:enthalpy}      
\centering          
\begin{tabular}{llc}
\hline\hline       
Species     & $\Delta H$ & Ref. \\
\hline                    
HNCO$^+$      & 243      & 1 \\
HOCN$^+$      & 274      & 2 \\
HCNO$^+$      & 292      & 2 \\
HONC$^+$      & 331.7    & 2 \\
HNOC$^+$      & 331.4    & 2 \\
H$_2$NCO$^+$  & 167      & 1 \\
HNCOH$^+$     & 183      & 3 \\
HCNOH$^+$     & 234.8    & 3 \\
H$_2$OCN$^+$  & 240.9    & 3 \\
H$_2$CNO$^+$  &   243    & 3 \\
\hline
\end{tabular}
\tablefoot{
(1) \citet{lias84}; 
(2) computed from \citet{luna96} using the experimental value for the most stable isomer HNCO$^+$; 
(3) computed from \citet{ijjaali01} using the experimental value for the most stable isomer H$_2$NCO$^+$.
}
\end{table}

\begin{table}
\caption{Assumed rate coefficients of the DR reaction of the CH$_2$NO$^+$ ions.}
\label{tab:DRrates}
\centering          
\begin{tabular}{lc}
\hline\hline       
Reaction & Rate Coefficient \\
         & ($\times$10$^{-7}$(T/300)$^{-0.5}$ cm$^3$ s$^{-1}$)  \\
\hline
${\rm  H_2NCO^+ + e \rightarrow HNCO +  H}$     &  2.50 \\
${\rm  H_2NCO^+ + e \rightarrow CO   +  NH_2}$  &  0.50 \\
${\rm  H_2NCO^+ + e \rightarrow NCO  +  H_2}$   &  1.50 \\
${\rm  H_2CNO^+ + e \rightarrow HCNO +  H}$     &  2.50 \\
${\rm  H_2CNO^+ + e \rightarrow CH_2 +  NO}$    &  0.50 \\
${\rm  HCNOH^+  + e \rightarrow HCNO +  H}$     &  2.50 \\
${\rm  HCNOH^+  + e \rightarrow HONC +  H}$     &  0.50 \\
${\rm  HCNOH^+  + e \rightarrow HCN  +  OH}$    &  0.50 \\
${\rm  HNCOH^+  + e \rightarrow HOCN +  H}$     &  0.50 \\
${\rm  HNCOH^+  + e \rightarrow HNCO +  H}$     &  2.50 \\
${\rm  HNCOH^+  + e \rightarrow HNC  +  OH}$    &  0.50 \\
${\rm  H_2OCN^+ + e \rightarrow HOCN +  H}$     &  1.50 \\
${\rm  H_2OCN^+ + e \rightarrow H_2O +  CN}$    &  0.50 \\
\hline
\end{tabular}
\end{table}

\end{appendix}

\end{document}